# A 0.6 Mpc HI Structure Associated with Stephan's Quintet

C.K. Xu[1,2], C. Cheng[1,2], P.N. Appleton[3], P.-A. Duc[4], Y. Gao[5,6], N.-Y. Tang[7], M. Yun[8], Y.S. Dai[1,2], J.-S. Huang[1,2], U. Lisenfeld[9,10], F. Renaud[11]

[1] Chinese Academy of Sciences South America Center for Astronomy, National Astronomical Observatories, CAS, Beijing 100101, People's Republic of China
[2] National Astronomical Observatories, Chinese Academy of Sciences (NAOC), 20A Datun Road, Chaoyang District, Beijing 100101, People's Republic of China
[3] Caltech/IPAC, MC 6-313, 1200 E. California Blvd., Pasadena, CA 91125, USA
[4] Université de Strasbourg, CNRS, Observatoire astronomique de Strasbourg, UMR 7550, F-67000 Strasbourg, France
[5] Department of Astronomy, Xiamen University, 422 Siming South Road, Xiamen 361005, People's Republic of China
[6] Purple Mountain Observatory & Key Laboratory for Radio Astronomy, Chinese Academy of Sciences, 10 Yuanhua Road, Nanjing 210023, People's Republic of China
[7] Department of Physics, Anhui Normal University, Wuhu, Anhui 241002, People's Republic of China
[8] Department of Astronomy, University of Massachusetts, Amherst, MA 01003, USA
[9] Dept. Física Teórica y del Cosmos, Campus de Fuentenueva, Edificio Mecenas, Universidad de Granada, E-18071 Granada, Spain
[10] Instituto Carlos I de Física Tórica y Computacional, Facultad de Ciencias, E-18071 Granada, Spain
[11] Department of Astronomy and Theoretical Physics, Lund Observatory, Box 43, SE-221 00 Lund, Sweden
Email: congxu@nao.cas.cn, chengcheng@nao.cas.cn

**Stephan's Quintet (SQ, distance=85±6 Mpc) is unique among compact groups of galaxies[1-12]. Observations have previously shown that interactions between multiple members, including a high-speed intruder galaxy currently colliding into the intragroup medium, have likely generated tidal debris in the form of multiple gaseous and stellar filaments[6,8,13], the formation of tidal dwarfs[7,14,15] and intragroup-medium starbursts[16], as well as widespread intergalactic shocked gas[5,10,11,17]. The details and timing of the interactions/collisions remain poorly understood because of the multiple nature[18,19]. Here we report atomic hydrogen (HI) observations in the vicinity of SQ with a smoothed sensitivity of $1\sigma=4.2\times10^{16}$ cm$^{-2}$ per channel ($\Delta v$=20 km s$^{-1}$; angular-resolution=4'), which are about two orders of magnitude deeper than previous observations[8,13,20,21]. The data reveal a large HI structure (linear scale ~0.6 Mpc) encompassing an extended source of size ~0.4 Mpc associated with the debris field and a curved diffuse feature of length ~0.5 Mpc attached to the south edge of the extended source. The diffuse feature was likely produced by tidal interactions in early stages of SQ (>1 Gyr ago), though it is not clear how the low density HI gas ($N_{HI} \lesssim 10^{18}$ cm$^{-2}$) can survive the ionization by the inter-galactic UV background on such a long time scale. Our observations require a re-**

thinking of gas in outer parts of galaxy groups and demand complex modeling of different phases of the intragroup medium in simulations of group formation.

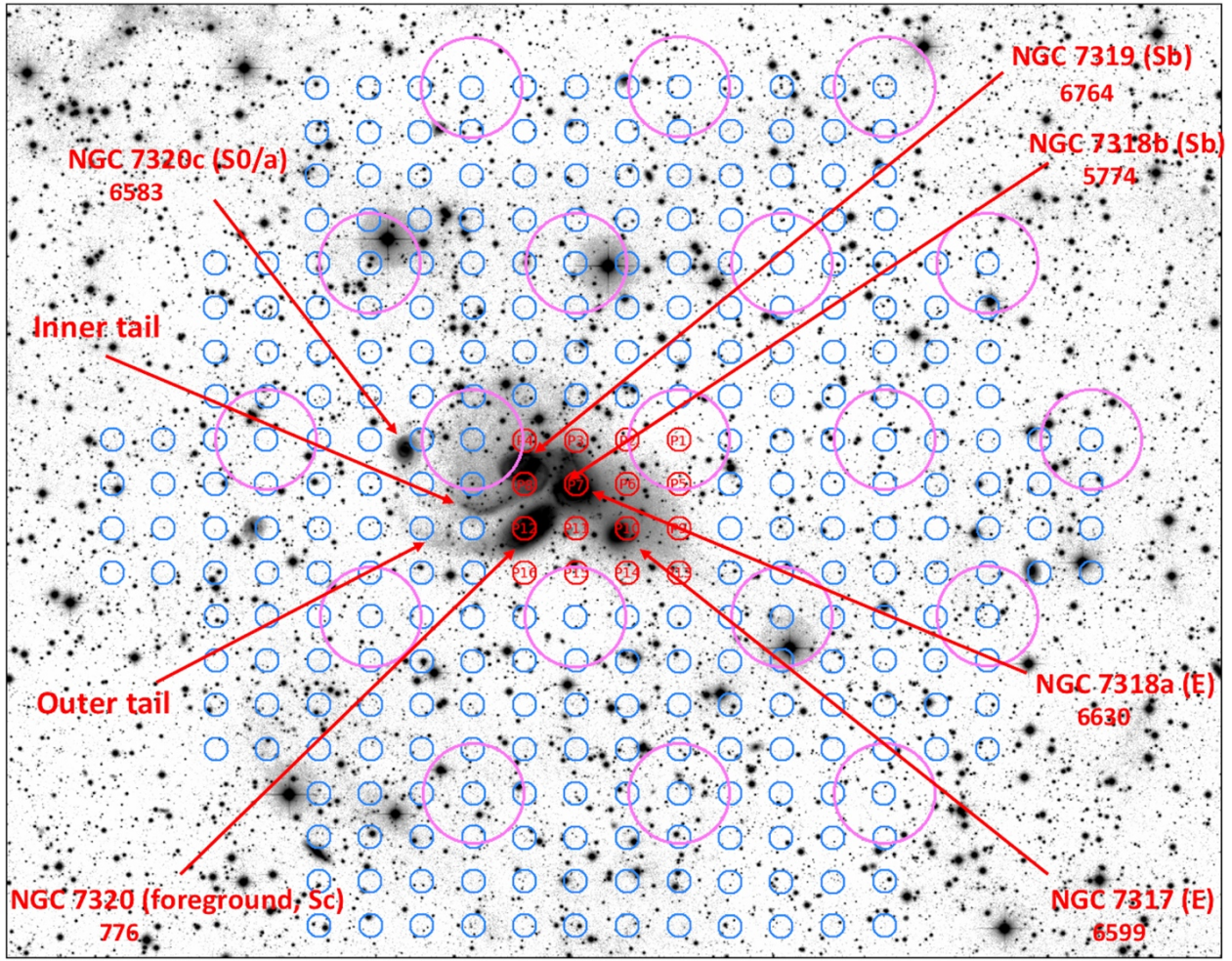

**Figure 1 | Member galaxies plus main tidal features in SQ and the sky coverage of the FAST observations of the SQ field.** The background image is the inverted greyscale map of the deep CFHT MegaCam r-band image[12]. The NGC name, radial velocity (in km s$^{-1}$), and the Hubble type are provided for each member galaxy in SQ. The core members include NGC 7317, NGC 7318a and NGC 7319, which are located near the group center and have similar radial velocities (6680±85 km s$^{-1}$). NGC 7320c is an "old intruder" which may be responsible for the formation of the inner and outer tails[6,19]. NGC 7318b is a "new intruder" currently colliding into the intragroup medium of SQ and triggering a large scale shock[5]. NGC 7320 is a foreground galaxy. The small blue circles mark the positions of individual beams in the HI mapping observations by the FAST 19-beam receiver. The observations were carried out at 16 slightly sparated pointings in a 4×4 rectangular grid. The small red circles mark the central positions of the FAST 19-beam receiver in these pointings, and the characters "*P1*", "*P2*", … "*P16*" inside the circles identify the different pointings. The large magenta circles (*D* = 2'.9, i.e. the half-power beam size) show the coverage of the 19 beams in the first pointing (see "FAST Observations" section in "Methods" for more details).

The HI is the least bound component of galaxies and is therefore the easiest (and hence first) to be stripped off and spread around during interactions. Thus, the distribution of the very diffuse HI and its velocity field can provide new information about the earliest interactions. In order to study the diffuse HI associated with SQ, we carried out deep mapping observations of the 21 cm HI emission over a region of ~ 30'× 30' centered on SQ (Figure 1) using the 19-beam receiver of Five-hundred-meter Aperture Spherical Telescope (FAST). The FAST observations and data reduction are described in Methods. As shown illustratively in Figure 2, the final data cube includes 304 spectra in $\Delta v$=20 km s$^{-1}$ channels covering the velocity range of 4600 – 7600 km s$^{-1}$, with an average rms of 0.16 mJy beam$^{-1}$ and an average beam size of 2'.9. The mapping satisfies the Nyquist sampling criterion with beams separated by 1'.4 in the RA direction and 1'.2 in the Dec direction. The original data cube has the HI column density sensitivity of 1σ=1.2×10$^{17}$ cm$^{-2}$ per channel, which is improved to 1σ=4.2×10$^{16}$ cm$^{-2}$ per channel when smoothed to 4'. Results of analysis of the whole data cube will be presented elsewhere. In this article we report only the discovery of a large diffuse structure in the velocity range of 6550 – 6750 km s$^{-1}$.

Panel a of Figure 3 presents the integrated HI emission map in the velocity range of 6550 – 6750 km s$^{-1}$ overlaid on the deep MegaCam optical color image[12]. The map has the angular resolution of 4'.0 and $\Delta v$=200 km s$^{-1}$ (ten times of the channel width) with an HI column density error of 1σ=1.34×10$^{17}$ cm$^{-2}$. The base contour starts from $N_{HI}$=7.4×10$^{17}$ cm$^{-2}$ (at 5.5-σ level). The map shows a large HI structure of ~0.6 Mpc in size, which has two parts: an extended source centered on SQ and a diffuse feature attached to the south edge of the source. The extended source encompasses the previously detected 6600 km s$^{-1}$ HI component associated with the debris field[8,13]. As shown by the cyan contours in Figure 3a, the high resolution (beam=19".4×18".6) VLA observations[8] detected only the high density part of the 6600 km s$^{-1}$ component ($N_{HI}$≥5.8×10$^{19}$ cm$^{-2}$) which is confined to the central region ($D \sim 0.1$ Mpc) of the extended source. Most of the high density HI gas traces the optically detected inner and outer tails[5] plus a compact cloud (on north-west of NGC 7319) coincident with the intragroup-medium starburst SQ-A[16]. Single dish HI mapping observations by Arecibo and GBT, which detected lower density HI gas at $N_{HI}$~ 5×10$^{18}$ cm$^{-2}$ albeit with lower angular resolutions (> 3'), have found evidence for this component to be extended on a scale of ~ 0.2 Mpc[20,21]. Our deeper FAST map shows an even larger size with a diameter of ~ 0.4 Mpc. The diffuse feature has a characteristic column density of ~ 7×10$^{17}$ cm$^{-2}$ in an elongated and curved structure of ~ 0.5 Mpc in length. It borders at the bottom of the FAST mapping and therefore may well reach beyond the map. The faint optical halo (the yellowish diffuse light around SQ in the optical color image) discovered previously[12] lies inside the extended source and has no spatial overlap with the newly discovered diffuse feature. The first moment map in Panel b of Figure 3 shows that in the velocity field the diffuse feature is linked smoothly with the extended source. The two green boxes marked by characters A and B in Panel a cover the entire diffuse HI feature. The sum of all spectra in these two boxes provides a good measure of the spectrum of the diffuse feature, which is presented in Panel c. The spectrum has a flux-density weighted mean velocity of 6633 km s$^{-1}$ and a rather narrow line width of $\Delta v_{20}$=160 km s$^{-1}$ (measured at 20% of the peak). The integrated flux is 0.42±0.03 Jy km s$^{-1}$, corresponding to an HI mass of 7.1±0.5×10$^8$ M$_\odot$ which is only ~3% of the total HI mass of SQ (2.45×10$^{10}$ M$_\odot$)[21]. It is worth noting that, although the GBT mapping observations found 65% more HI than the VLA observations, SQ is still slightly deficient in HI abundance compared to normal galaxies (by a factor of ~1.3)[4,21]. The very diffuse HI ($N_{HI}$<3×10$^{18}$ cm$^{-2}$) discovered in this work does not change this HI deficiency significantly.

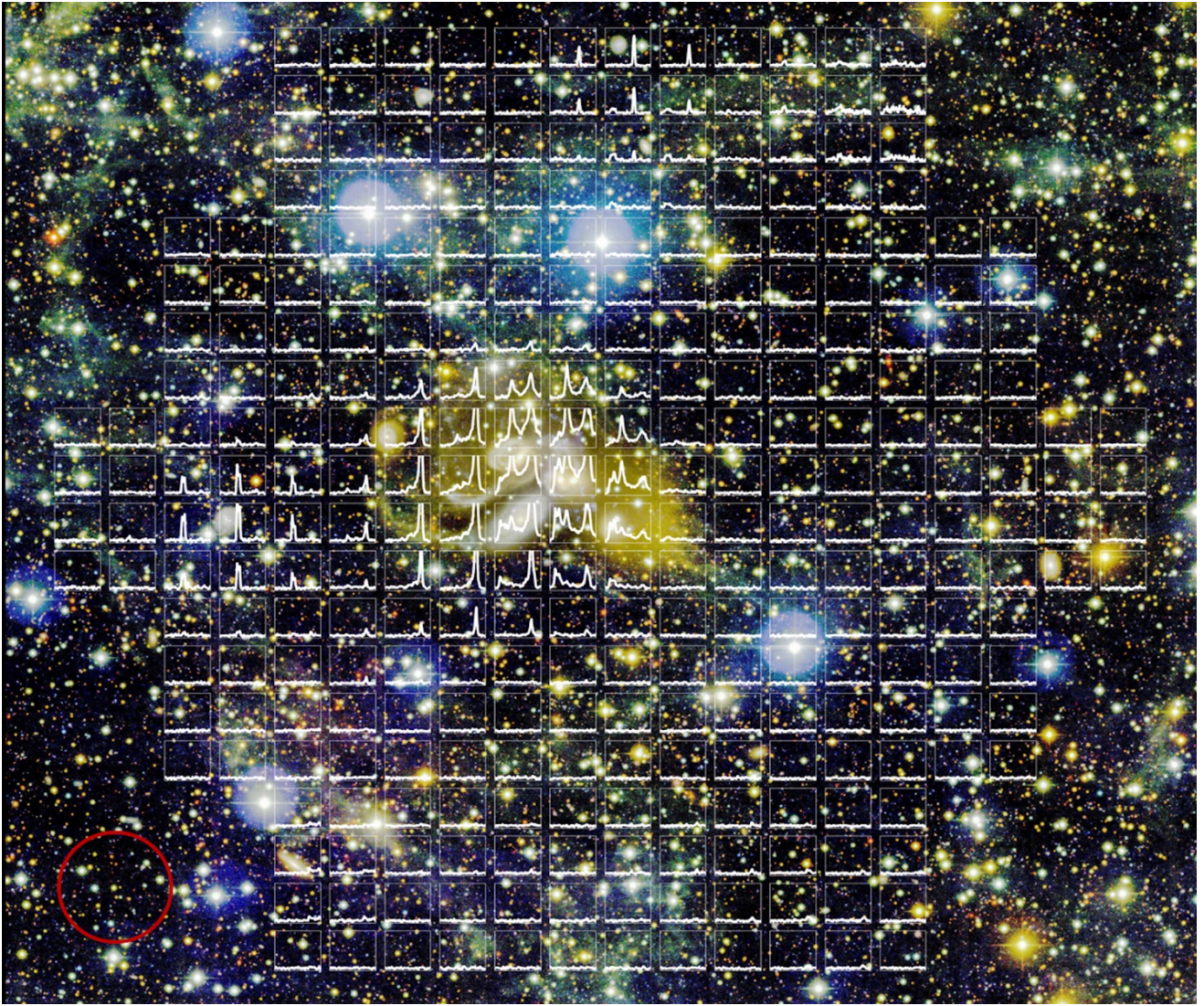

**Figure 2. | Illustrative plot of the 304 spectra of the HI emission in the SQ field.** The underlying optical color image (*u, g, r*) is obtained in the deep CFHT MegaCam observation with limiting surface brightness of 29.0, 28.6 and 27.6 mag arcsec$^{-2}$ for the three bands, respectively[12]. The spectra cover the velocity range of 4600 – 7600 km s$^{-1}$ with an average rms of 0.16 mJy beam$^{-1}$ per channel (Δv=20 km s$^{-1}$). The center of each spectrum coincides with the pointing position of the FAST beam with which it was obtained. The final data cube includes 304 spectra in 20 km s$^{-1}$ bins. The red circle in the bottom-left corner shows the size of the FAST beam (2'.9, or 72.5 kpc in linear scale).

Two new detections of unresolved sources can also be found in Figure 3a. NGC 7320a, detected with S/N=36, has an HI mass of $M_{HI}$=6.3±0.2×10$^8$ M$_\odot$ and a $v_{HI}$=6702±24 km s$^{-1}$. The other source Anon 7, a 4.4σ detection, has an HI velocity of 6654±16 km s$^{-1}$ and an HI mass of 2.2±0.5×10$^8$ M$_\odot$. More discussions about these two sources are given in Methods.

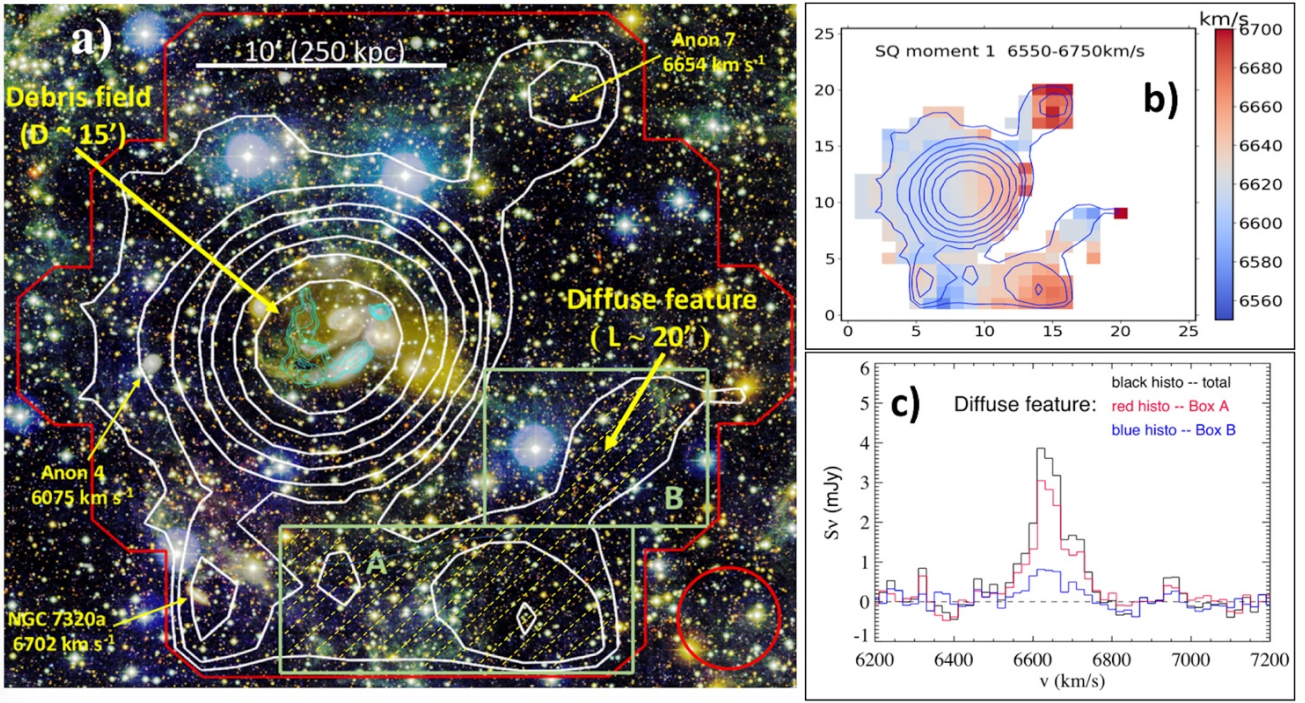

**Figure 3. | The HI emission in the velocity range of 6550 – 6750 km s$^{-1}$. a):** Contour map of the integrated HI emission in the velocity range of 6550 – 6750 km s$^{-1}$ overlaid on the composite color image of the deep CFHT MegaCam observation. The red circle at bottom-right illustrates the angular resolution of the FAST map after smoothing (half-power-beam-width=4'.0). The contours start from $N_{HI}$ =7.4×10$^{17}$ cm$^{-2}$ (at 5.5-σ level) with an increment of a factor of 2. The red lines delineate the boundary of the FAST observations. The cyan contours in the center are adopted from the VLA observations for the 6600 component of SQ$^8$, with angular resolution of 19".4×18".6. They have the base level at $N_{HI}$ =5.8×10$^{19}$ cm$^{-2}$ and the increment of a factor of 2. The area occupied by the newly discovered diffuse feature is filled with a hatch pattern consisting of thin yellow dotted lines. The two green boxes marked by characters A and B cover the diffuse feature. **b):** False color map of the velocity field of the HI emission in the velocity range of 6550 – 6750 km s$^{-1}$ overlaid by the same contour map shown in panel a). The units of the *x* and *y* axes are in pixels (pixel size: 1'.4×1'.2). **c):** The HI spectrum of the diffuse feature. The red line is the summation of all spectra inside Box A, and the blue line the summation of those inside Box B. The black line is the sum of the two spectra of Box A and Box B.

We examine in Figure 4 the individual spectra in Box A and Box B in order to investigate the physical nature of the diffuse feature. No diffuse stellar radiation is detected in these regions down to the limit of the deep MegaCam image. Spectra of beams with detections of S/N>4 are marked by pink boxes and those with 3< S/N⩽4 by green boxes. Marked in Figure 4 are also galaxies brighter than r = 20 mag found in the SDSS photo-z catalog$^{22}$. Only two of them have photo-z <0.1 (marked by red circles) and the remaining 28 have photo-z⩾0.1 (orange circles). Given the 1-σ error of the photo-z (δz/(1+z) = 0.02)$^{22}$ and the redshift of SQ (z=0.02), galaxies with photo-z⩾0.1 are very unlikely to be at the same redshift of SQ (probability < 0.001). We can rule out with high confidence the possibility of the diffuse feature being associated with a collection of gas-rich galaxies (even including those fainter than *r* = 20 mag), because it needs at least four such galaxies to cover all beams with significant detections of S/N>4 (one for those in the upper-left corner of Box A, two for those in the right half of Box A, and one for that in Box B) and the probability that they happen to have about the same radial velocity is extremely low.

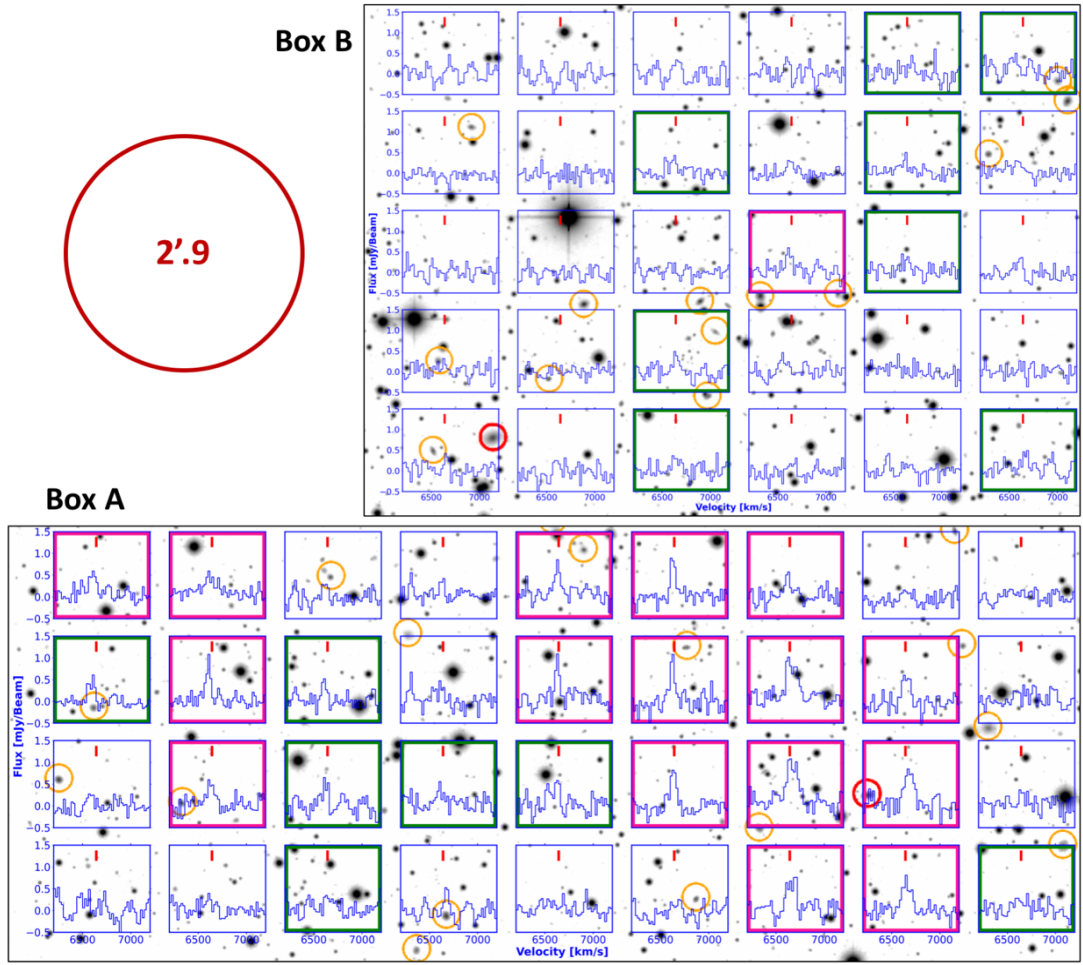

**Figure 4 | Individual spectra in the diffuse feature region.** Individual spectra in Box A and Box B in Figure 3a are overlaid on the inverted greyscale map of CFHT MegaCam r-band image. The center of each spectrum coincides with the pointing position of the FAST beam with which it was obtained. All spectra are plotted with the same velocity range of [6200, 7200] km s$^{-1}$ and flux density range of [-0.5, 1.5] mJy beam$^{-1}$. The short vertical red line on top of every spectrum marks the position of *v*=6642 km s$^{-1}$ (the flux-density-weighted mean velocity of the diffuse feature). The spectra with detections of S/N > 4 are identified by pink boxes and those with detections of 3 < S/N ≤ 4 by green boxes. Galaxies brighter than *r*=20 mag are marked by red circles if they have photo-z < 0.1 or orange circles if they have photo-z≥0.1. The large red circle in upper-left corner (outside the boxes) illustrates the half-power beam size of the FAST observations.

HI clouds without stellar counterparts have been found in and around many galaxy groups/clusters[23-26]. Most of them, as the authors of Ref. 27 argued, can be explained by tidal debris of galaxy interaction involving very extended HI disks instead of "dark" or "almost dark" galaxies. Given its location and velocity, the diffuse feature is most likely to be related to the debris field. A hypothetical scenario for the formation of the diffuse feature is that NGC 7320a ($v$=6702 km s$^{-1}$ and currently ~ 300 kpc away from the SQ center) passed through the SQ center ~1.5 Gyr ago (assuming a relative transverse velocity of 200 km s$^{-1}$) and pulled out from one of the core member galaxies of SQ a tidal tail which developed into the diffuse feature by now. Another possibility is that, like the large Leo Ring ($D$ ~ 0.25 Mpc)[28], the diffuse feature could be the product of a high-speed head-on collision between another old intruder and one of the core members of SQ. In this scenario, the collision triggers an expanding density wave which pushes gas in an extended HI disk of the target galaxy outward to form a very large ring, of which the diffuse feature is the high-density part. A candidate of such an intruder could be Anon 4 ($v$=6057 km s$^{-1}$, $M_{HI}$=1.1×10$^9$ M$_\odot$)[8] which spatially coincides with optical galaxy LEDA 141041 ($B$=18.4 mag). It has a relative radial velocity of ~ 600 km s$^{-1}$ and a projected distance of ~ 0.2 Mpc from the SQ center. If the relative transverse velocity is ~200 km s$^{-1}$, it would have taken ~1 Gyr for Anon 4 to move to the current position after the collision. Both scenarios proposed above suggest a formation time of the diffuse feature >1 Gyr ago. They both are based on analogies to cases studied in simulations in the literature which demonstrate that diffuse HI features without stellar component can be produced in galaxy-galaxy interations[27,28]. However two questions must be answered: (1) Can the tidal feature in either of the scenarios survive the subsequent interactions that triggered the formation of the inner and outer tails of SQ about 3 – 8×10$^8$ years ago[18,19]? (2) Can HI structures with column density as low as $N_{HI} \lesssim 10^{18}$ cm$^{-2}$ exist on time scales of ~1 Gyr? These questions can only be answered by more sophisticated models that are built upon the existing simulations for the formation and evolution of SQ[18,19]. It has been argued that cold gas of $N_{HI} \leq 2\times10^{19}$ cm$^{-2}$ cannot stay neutral in the intergalactic UV background for more than 500 Myr[21,29]. A plausible solution for this problem is the physical mechanism involving the transition between ionized and neutral phases due to thermal instabilities in the low density gas[30,31]. New simulations, which are beyond the scope of this article, shall explore this mechanism.

## Methods

**FAST Observations:**

The deep HI mapping observations were carried out in September and October of 2021 using the FAST 19-beam receiver in the standard ON-OFF mode with the total observation time of 22.4 hours including overheads (Extended Data Table 1). The FAST 19-beam L-band Array is currently the largest multi-beam feed array for HI observations in the world. Details about its properties and performance can be found in Ref 32. The 19 beams are arranged in a hexagonal configuration with the neighboring beams separated by 5'.7. The observations have the central frequency of 1391.64 MHz and the frequency coverage of 1050 – 1450 MHz with a resolution of 7.63 kHz ($\Delta v$ = 1.65 km s$^{-1}$). For the 19 beams, the average half-power beam-width at 1391 MHz is 2'.9 (Extended Data Table 2). In order to meet the Nyquist sampling criterion and fill the gaps between beams in the focal plane, we did 16 pointings in a 4×4 rectangular grid in the north-up orientation (Figure 1). The final mapping covers a region of ~ 30'× 30' centered on SQ with 304 sky pixels (beam positions in the sky), and the separation between the nearest pixels is 1'.4 in the RA direction and 1'.2 in the Dec direction. The 1-σ pointing error of individual beams is 7''.9 [32]. At each pointing, 6 cycles of ON-OFF integrations were conducted, with the OFF position at 40' south-east from the ON position. Each ON or OFF took 300s integration with the sampling frequency of 1 Hz. The total on-target time for each pixel is 1800s (Extended Data Table 1). In order to minimize the effects of standing waves and sidelobes, all observations were confined within the zenith angle of 20 degrees.

Compared to the scan-mapping mode which is more suitable for large sky surveys such as the FAST Extragalactic HI Survey (FEHIS)[33], our observational strategy provides an alternative for deep and small maps (≤ 30') which can take the advantage of the ON-OFF mode in more accurately removing various systematic effects such as the standing waves and baseline wobbling.

**Data Reduction and Calibration:**

For each sky pixel observed by a given beam, we reduced the spectral data following a similar procedure as presented in Ref 34. The spectra of the two polarizations were reduced separately (and eventually combined after the consistency check). The data were grouped in ON-OFF cycles. Each ON (or OFF) has 300 samplings. They were averaged and calibrated (i.e. converted from ADU to Kelvin), resulting in a single raw spectrum. During the observations, a calibration signal (CAL) of 10 K was injected for a duration of 20s at the beginning of every ON-OFF cycle, and these data were used to calibrate the antenna temperature $T_a$. The calibration error is on the order of 10%. Repeating this, we obtained a raw spectrum for every ON or OFF. As examples, the upper panel of Extended Data Figure 1 presents the individual raw ON – OFF spectra obtained by M01 beam during the first pointing observation (the *P1* pixel in Figure 1). The mean of these spectra is presented in the middle panel of Extended Data Figure 1. It is affected significantly by the standing waves, which can be well fitted by a sine function locally. The spectrum has been converted from $T_a$ (in K) to flux density (in mJy). The gain factor that converts $T_a$ to flux density (in the units of K/Jy) depends on the frequency and varies from beam to beam. The values for individual beams at 1391 MHz, derived by interpolating values at other frequencies adopted from Ref 32, are presented in Extended Data Table 2. The next step was to remove the standing waves together with the baseline from the spectrum. The bottom panel of Extended Data Figure 1 presents the final spectrum at sky pixel *P1* after the subtraction of a baseline modelled by a sinusoidal (representing the standing waves) plus a polynomial (for the baseline gradient). We converted the frequency to velocity by adopting the optical redshift convention and the local standard of rest (LSR) reference frame, and re-binned the spectrum into bins of $\Delta v = 20$ km s$^{-1}$. The above process was carried out repeatedly for every sky-pixel observed in our observations.

It is worth noting that, if the standing waves and baseline vary significantly during a given observation, it is better to do the standing-waves-and-baseline removal for spectra of individual ON-OFF's instead of doing it for their mean. However, in a test in which we did the standing waves and baseline subtraction for each ON-OFF and then used the median of the baseline-subtracted spectra of ON-OFF's as the final spectrum at a given pixel, we got a noisier product. It appears that the wavelength of the standing waves (in the frequency domain) in a given observation is rather constant (though the phase changes from cycle to cycle). Consequently, the effect of the standing waves in the mean of the spectra of individual ON-OFF's is still a well-defined sinusoidal with the same wavelength which can be easily removed. Hence, because the mean spectrum is less noisy than individual spectra and therefore a more accurate model for the standing waves and the baseline can be obtained, subtracting the baseline model from the mean spectrum can achieve a better result.

The final data cube was constructed from the 304 individual spectra so obtained, with velocity coverage of 4600 – 7600 km s$^{-1}$ in 20 km s$^{-1}$ bins. A uniform half-power beam-width of 2'.9 was adopted for all spectra, neglecting the small variation of the beam size among different beams. The actual half-power beam-width of individual beams at 1391 MHz, derived from interpolations of values adopted from Ref 2, are listed in Extended Data Table 2. The units of the flux density of the spectra in the cube are in mJy beam$^{-1}$. To find the flux density in the units of mJy for a given spectrum, a factor of $A=(B/2'.9)^2$ should be multiplied to the value taken from the cube, where $B$ is the half-power beam-width (in arcmin) of the beam with which the spectrum was observed.

In the 16-pointing observations, each of the 19 beams covered 16 adjacent sky pixels. The mean and the standard deviation of the measured rms noise of these 16 spectra are also listed in Extended Data Table 2. The rms noise of each spectrum was measured in the two velocity intervals of 4700 – 5000 km s$^{-1}$ and 7000 – 7500 km s$^{-1}$, where no HI signal was detected. Beam M16 stands out as the noisiest beam in the array, with a mean rms of 0.26 mJy beam$^{-1}$ and a standard deviation of 0.14 mJy beam$^{-1}$. The false color map of the rms noise in the left panel of Extended Data Figure 2 shows that indeed pixels in the top-right corner which were covered by beam M16 have higher noise than others. The histogram of the distribution of the rms noise at all sky pixels is shown in the right panel of Extended Data Figure 2. The mean of the rms is 0.16 mJy beam$^{-1}$ with a standard deviation of 0.05 mJy beam$^{-1}$. It is worth noting that strong radio frequency interference (RFI) in the frequency range of our observations was a serious issue in the early stage of this project. The operation team of FAST did an excellent work in discovering and removing the source of the RFI in a relatively short time. All of our observations were carried out after the removing of the RFI source. Consequently, our observations were not affected by any significant RFI.

**Smoothing:**
The HI data cube obtained above is highly redundant in the sense that a sky area of the size of a single beam ($D=2'.9$) is covered by multiple beams (beam-separation: $1'.4\times1'.2$). When making channel maps and integrated emission maps from the data cube, applying a Gaussian-kernel convolution (i.e. smoothing) makes good use of this redundancy. This minimizes the noise due to the signal fluctuations in adjacent beams and results in significant improvement in the HI column-density sensitivity. The only cost is a slight degradation in the angular resolution. For single channel maps, the mean rms of 0.16 mJy beam$^{-1}$ corresponds to a HI column density sensitivity of $1\sigma = 1.2 \times 10^{17}$ cm$^{-2}$ per channel ($\Delta v=20$ km s$^{-1}$). When a smoothing with a Gaussian kernel of Full-Width-at-Half-Maximum (FWHM) = $2'.8$ is applied, the HI column density sensitivity is improved by a factor of 2.9 to $1\sigma = 4.2\times10^{16}$ cm$^{-2}$ per channel while the angular resolution is degraded only slightly (by a factor 1.4) to $4'.0$. The improvement in the HI column density sensitivity is particularly important for the exploration of diffuse extended emission. In Extended Data Figure 3 we present the contour map of the integrated HI emission in the velocity range of 6550 – 6750 km s$^{-1}$ before the smoothing. Compared to the map after the smoothing (Figure 3a), the low HI column density features in Extended Data Figure 3 are more fragmented mainly due to the signal fluctuations in adjacent beams. In the linear scale the pre- and after-smoothing resolutions are 72.5 kpc and 100 kpc, respectively, not a significant difference given that we are searching for extended diffuse HI gas of linear scale of a few 100 kpc.

**Sidelobe Correction:**
The data cube is corrected for the sidelobes utilizing the images of individual beams of the 19-beam receiver[32], which provide information of the point spread functions (PSF's) of the beams. For each beam, a "sidelobe responsivity function" is defined by the difference between the PSF and the "main beam", the latter is approximated by a 2-D Gaussian with the FWHM equal to the half-power beam-width. In the calculation of the sidelobe corrections, we consider only the effects due to the central extended source associated with the SQ group. The FAST observations also detected numerous other HI sources in the SQ neighborhood in the velocity range of 5500 – 7000 km s$^{-1}$. They are much fainter than the central SQ source and therefore their contributions to the sidelobes are neglected.

The first step is to estimate the sidelobe contribution to the map of integrated HI emission in the velocity range of 6550 – 6750 km s$^{-1}$, which encompasses the peak of the HI spectrum of SQ$^{21}$. The original integrated HI emission map (`observed map') is firstly deconvolved with the main beam of M01 (the central beam of 19-beam receiver). Then all pixels outside a circular aperture of $D$=10'.6, within which the central extended source is located, is masked. The result is then taken as our approximation for the `truth map'. The sidelobe contribution to any given pixel in the observed map is estimated by the following equation: $F_{sidelobe}(x_i, y_j) = \Sigma_m \Sigma_n T(x_m, y_n) \times R_k(x_m-x_i, y_n-y_j)$, where $x_i$ and $y_j$ are the coordinates of the pixel center, $T(x_m, y_n)$ is the flux in the truth map in the pixel at $x_m$ and $y_n$, $R_k$ is the sidelobe responsivity function of the beam that is pointed at the pixel $(x_i, y_j)$, and $\Sigma_m$ and $\Sigma_n$ are summations along the $x$ and $y$ directions, respectively. Extended Data Figure 4 presents the map of the sidelobe contribution so estimated overlaid by contours of the map of integrated HI emission in the velocity range of 6550 – 6750 km s$^{-1}$ (without the sidelobe correction, smoothed by a Gaussian kernel of FWHM=2'.8). It shows that the sidelobes contribute significantly at the edge of the debris field but have minimal effect to the diffuse feature in the south.

Neglecting the frequency dependences of the shapes of both the sidelobes and the central SQ source, we estimate the sidelobe contribution to each channel in the data cube by scaling the map in Extended Data Figure 4 with a factor of $C_v = S_v / S_{6550-6750}$, where $S_v$ (in mJy) is the flux density of SQ in the given channel, and $S_{6550-6750}$ (in mJy km s$^{-1}$) is the integrated flux of SQ in the velocity range of 6550 – 6750 km s$^{-1}$. Finally, we obtain the sidelobe corrected data cube by subtracting the sidelobe contribution so estimated from every channel map in the cube. The 304 spectra in the resulting data cube are presented illustratively in Figure 2.

**Detections of Two New HI Sources in the SQ Neighborhood:**
In the velocity range of 6550 – 6750 km s$^{-1}$ we detected two new unresolved HI sources in the SQ neighborhood (Extended Data Table 3). Their HI spectra are presented in Extended Data Figure 5. The spectrum of the source associated with NGC 7320a shows a typical double-horn profile consistent with the highly edge-on optical morphology of the galaxy. In order to confirm the association of the HI source and the optical galaxy, we made a long-slit optical spectroscopic observation (1 hour exposure) for NGC 7320a in the night of Dec. 21, 2021 using the 2.4 meter telescope at Lijiang Observatory. The optical spectrum is presented in Extended Data Figure 6. The radial velocity obtained from the optical spectrum is 6729±59 km s$^{-1}$, consistent with the HI velocity (6702±24 km s$^{-1}$). The other source is a 4.4σ detection without obvious optical counterpart, and therefore we name it Anon 7 following the convention in the literature[8,13].

**Data availability**
Data availability: Observational data are available from the FAST archive (http://fast.bao.ac.cn) 1 year after data collection, following FAST data policy. The data that support the findings of this study are openly available in Science Data Bank at https://www.scidb.cn/s/jiIfee

**Code availability**
The Python and IDL code for the data reduction pipeline is available at https://www.scidb.cn/s/jiIfee

**Acknowledgements:**
This work is supported by the National Key R&D Program of China No. 2017YFA0402704 and National Natural Science Foundation of China (NSFC) No. 11873055 and sponsored (in part) by the Chinese Academy of Sciences (CAS) through a grant to the CAS South America Center for Astronomy (CASSACA). C.K.X. acknowledges NSFC grant No. 11733006. C.C. acknowledges NSFC grant No. 11803044 and 12173045. N.-Y.T. acknowledges support from the Cultivation Project for FAST Scientific Payoff and Research Achievement of CAMS-CAS and Natural Science Foundation of China grants (NSFC No. 11803051). J.-S.H. acknowledges NSFC grant No. 11933003. UL acknowledges support from project PID2020-114414GB-100, financed by MCIN/AEI/10.13039/501100011033, from project P20_00334 financed by the *Junta de Andalucia and from FEDER/Junta de Andalucía-Consejería de Transformaciòn Econòmica, Industria, Conocimiento y Universidades/Proyecto A-FQM-510-UGR20*. F. R. acknowledges support from the Knut and Alice Wallenberg Foundation. This work made use of data from FAST, a Chinese national mega-science facility built and operated by the National Astronomical Observatories, Chinese Academy of Sciences. We thank Peng Jiang, Ligang Hou, Chun Sun and other FAST operation team members for supports in the observations and data reductions, and Hai-Cheng Feng, Yongping Huang for helping the optical spectroscopic observation of NGC7320a. Support of the staff of the Lijiang 2.4 m telescope is acknowledged. Funding for the Lijiang 2.4 m telescope has been provided by the Chinese Academy of Sciences and the People's Government of Yunnan Province. This research has made use of the NASA/IPAC Extragalactic Database (NED), which is operated by the Jet Propulsion Laboratory, California Institute of Technology, under contract with the National Aeronautics and Space Administration. We dedicate this article to the memory of Dr. Yu Gao, a coauthor of the article who passed away recently.

**Author contributions** C.K.X.: observing, data reduction, calibration, modelling, interpretation. C.C.: data reduction, calibration, modelling, interpretation, ancillary data. P.N.A.: interpretation, proof reading. P.-A.D., M.Y.: interpretation, ancillary data. N.-Y.T.: observing, calibration. Y.G., Y.S.D., J.-S.H., U.L., F.R.: interpretation.

**Competing interests:** The authors declare no competing interests.

**Correspondence and requests for materials** should be addressed to C.K. Xu and C. Cheng.

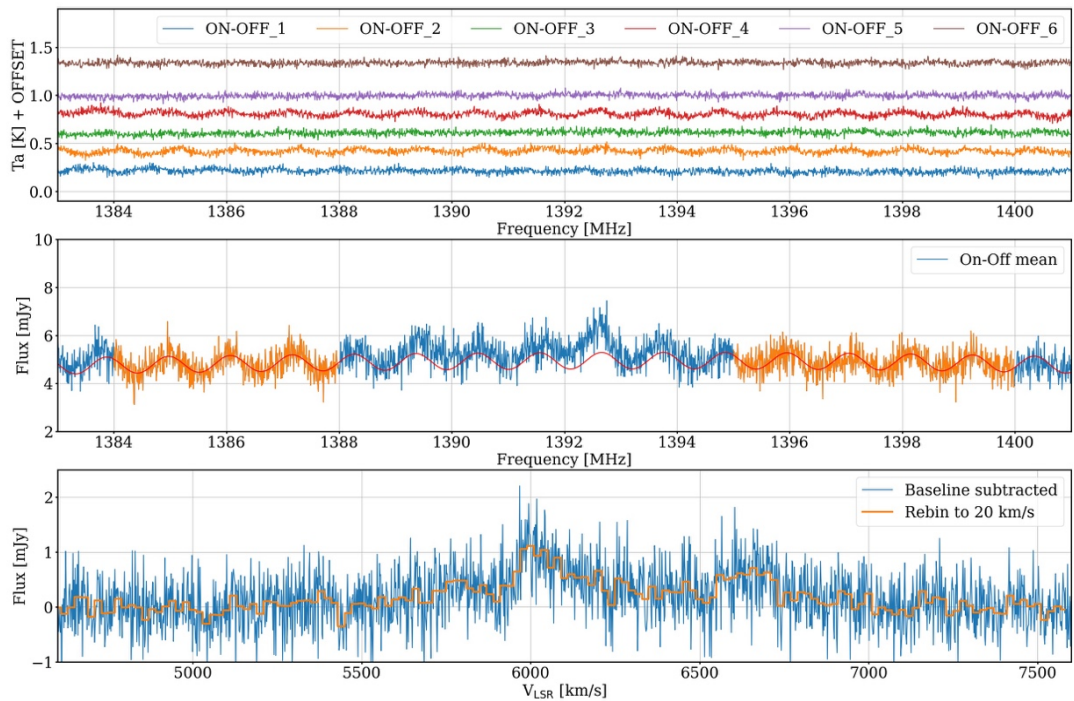

**Extended Data Figure 1 | Examples of intermediate products of spectral data reduction. Upper Panel:** Individual ON-OFF spectra obtained by beam M01 in Pointing 1 observation (i.e. the data obtained at the pixel *P1* in Figure 1). An offset of 0.2 K is added to every spectrum relative to the previous one in order to make them separated from each other. **Middle Panel:** The mean of the ON-OFF spectra in the upper panel. The units of the flux density are converted from K to mJy using the gain factor taken from Extended Data Table 2. The red line shows the baseline model derived by fitting the parts of the spectrum free of signals (marked by orange color) using a sinusoidal (representing standing waves) plus a polynomial (for baseline gradient). **Bottom Panel:** The spectrum after the baseline removal. The frequency is converted to radial velocity in the optical convention relative to the local standard rest reference frame (LSR). The orange line presents the rebinned spectrum with the velocity bin width of 20 km s$^{-1}$.

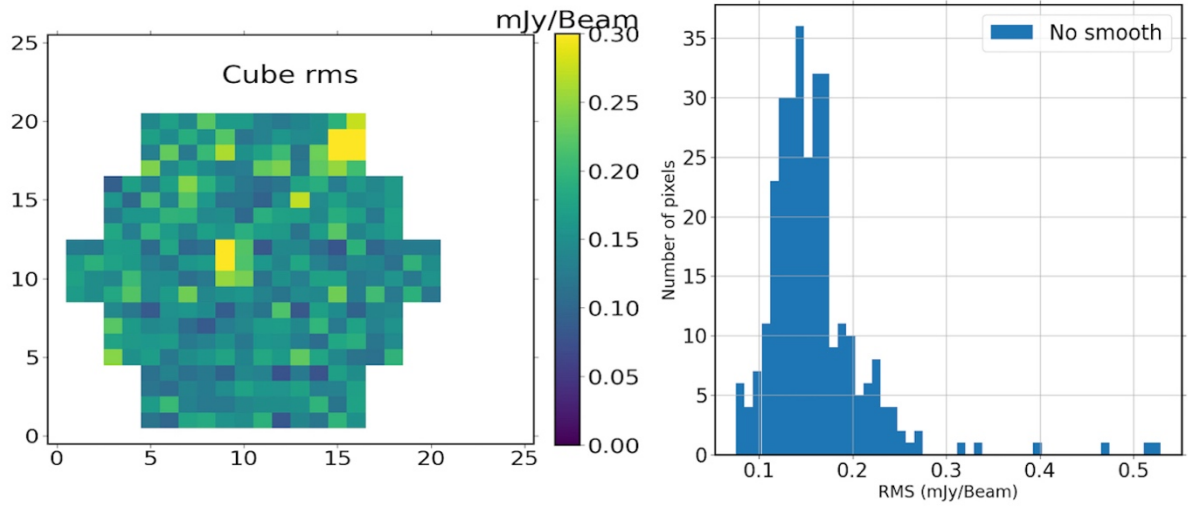

**Extended Data Figure 2 | Spectral RMS noise of individual sky pixels. Left:** The false color image of the rms noise at different sky pixels. **Right:** Histogram of the distribution of the rms noise.

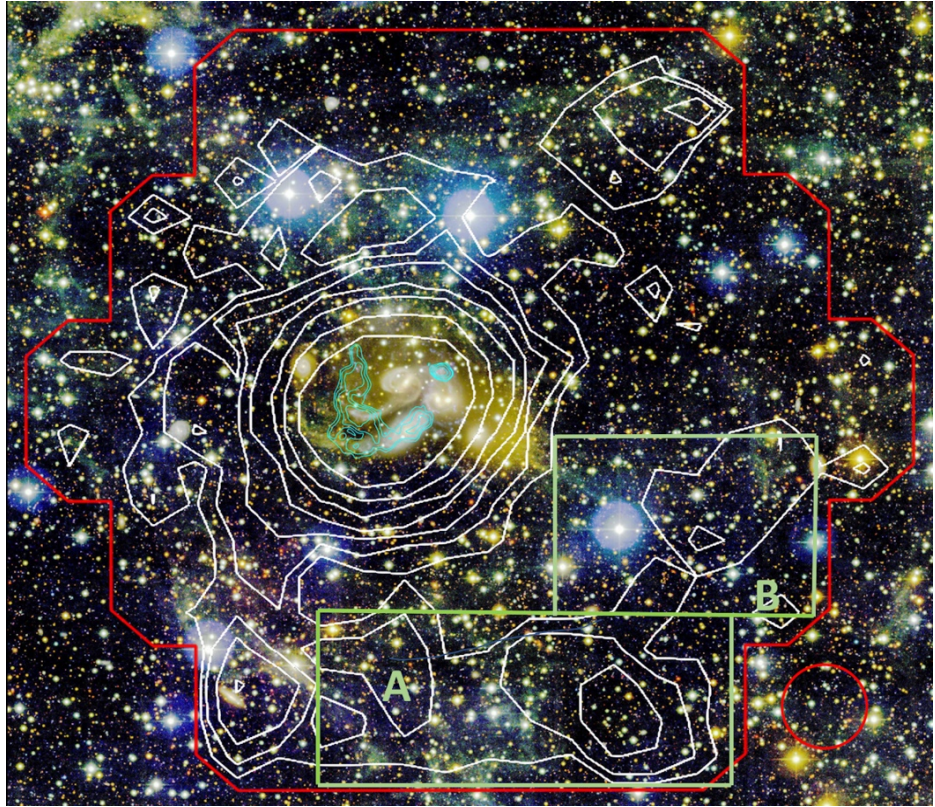

**Extended Data Figure 3 | The HI emission in the velocity range of 6550 – 6750 km s$^{-1}$ (unsmoothed).** Contour map of integrated HI emission (unsmoothed) in the velocity range of 6550 – 6750 km s$^{-1}$ overlaid on the color image of the deep CFHT MegaCam observation. The red circle at bottom-right illustrates the angular resolution (2'.9) of the FAST map (unsmoothed). The contours start from $N_{HI}$ =7.4 × 10$^{17}$ cm$^{-2}$ (at 2-σ level) with an increment of a factor of 2. The red lines delineate the boundary of the FAST observations. The cyan contours in the center are adopted from the VLA observations for the 6600 component of SQ[8], with angular resolution of 19".4×18".6. They have the base level at $N_{HI}$ =5.8×10$^{19}$ cm$^{-2}$ and the increment of a factor of 2. The same Box A and Box B that mark the location of the diffuse feature in Figure 3a are plotted here to facilitate the comparison between the two figures.

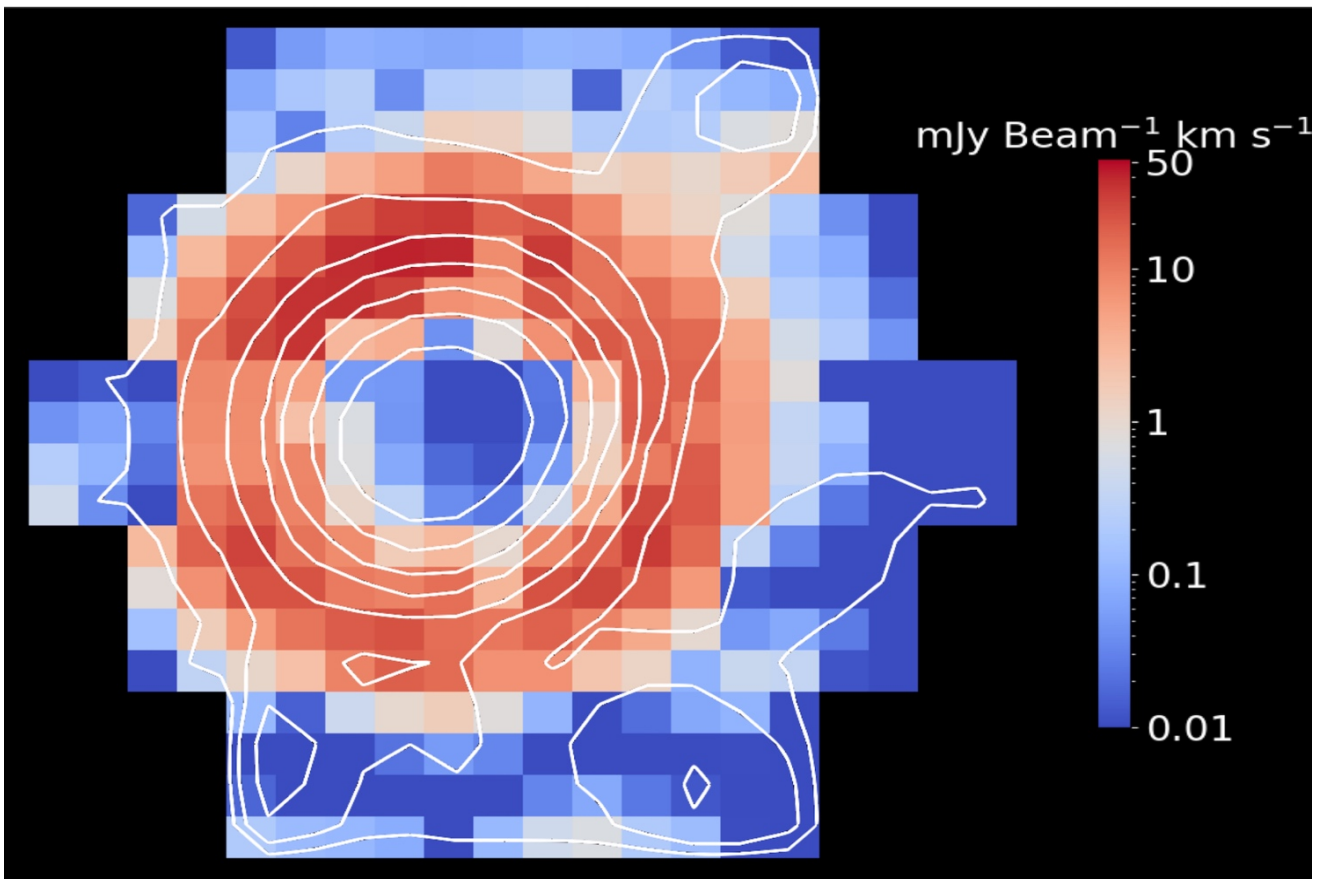

**Extended Data Figure 4 | Sidelobe contribution to the integrated HI emission in the velocity range of 6550 – 6750 km s$^{-1}$.** False color image of the sidelobe contribution overlaid by the contour map of integrated HI emission in the velocity range of 6550 – 6750 km s$^{-1}$ (before the sidelobe correction and after the convolution by a Gaussian kernel of FWHM=2'.8). The contours start from 20 mJy km s$^{-1}$ beam$^{-1}$ (corresponding to $N_{HI}$ =7.4 × 10$^{17}$ cm$^{-2}$ for the original beam of FWHM=2'.9) with an increment of a factor of 2.

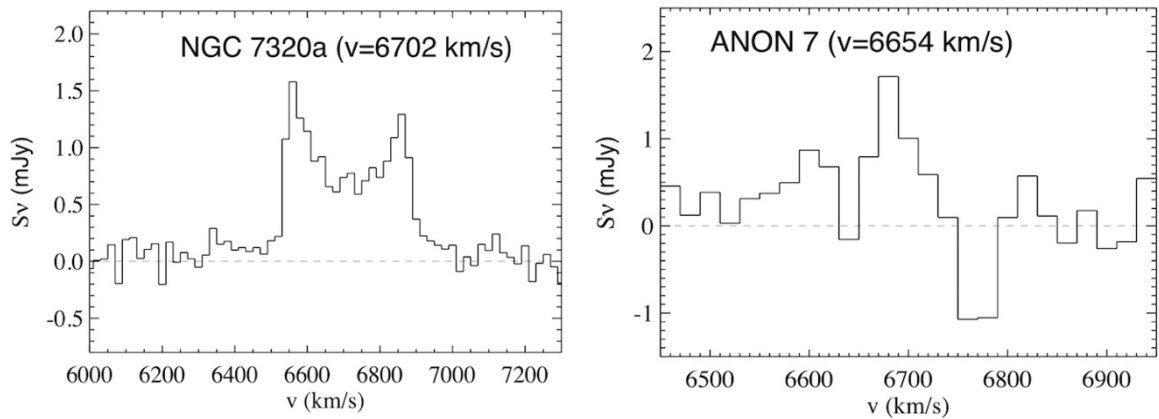

**Extended Data Figure 5 | HI spectra of NGC 7320a and Anon 7.** HI spectra of two newly detected HI sources (NGC 7320a and Anon 7) in the SQ neighborhood.

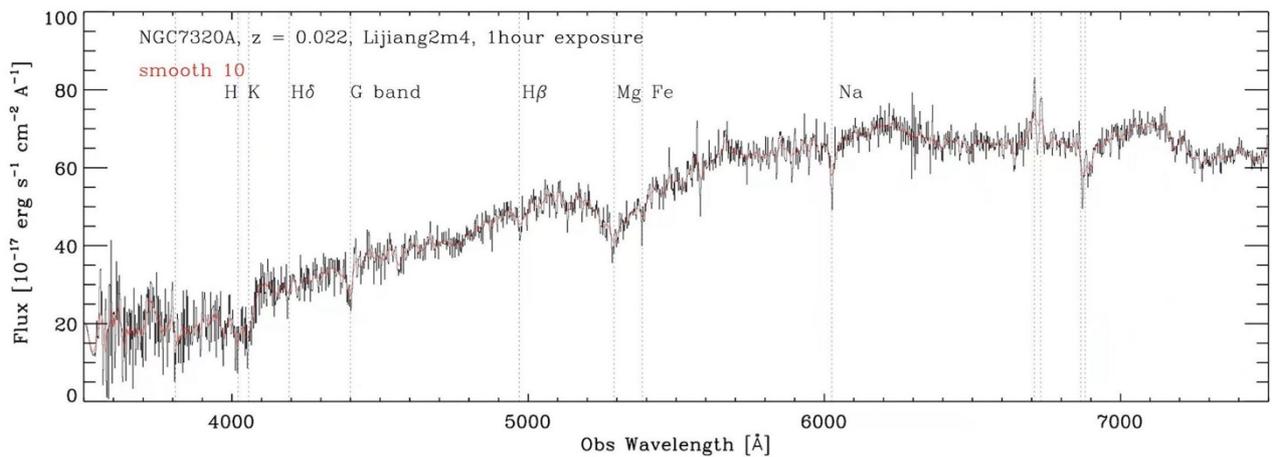

**Extended Data Figure 6 | Optical spectrum of NGC 7320a.** Optical spectrum of NGC 7320a obtained at Lijiang 2.4 meter telescope. The vertical dotted lines mark the positions of corresponding emission/absorption lines redshifted according to the best-fit redshift $z$=0.02243 ($v$=6729 km s$^{-1}$).

# Extended Data Table 1 | FAST Observations

| Obs. | date | $t_{total}$ [a] (secs) | R.A. [b] (2000J) | Decl. [b] (2000J) | cent freq. (MHz) | ON-OFF cycles | $t_{on}$ [c] (secs) |
|---|---|---|---|---|---|---|---|
| Pointing 1 [d] | 2021-09-09 | 5580 | 22:35:43.37 | +33:59:20.6 | 1391.64 | 6 | 1800 |
| Pointing 2 [d] | 2021-09-14 | 5580 | 22:35:50.39 | +33:59:20.6 | 1391.64 | 6 | 1800 |
| Pointing 3 [d] | 2021-09-11 | 5580 | 22:35:57.31 | +33:59:20.6 | 1391.64 | 6 | 1800 |
| Pointing 4 [d] | 2021-09-13 | 5580 | 22:36:04.23 | +33:59:20.6 | 1391.64 | 6 | 1800 |
| Pointing 5 | 2021-09-15 | 4860 | 22:35:43.37 | +33:58:06.1 | 1391.64 | 6 | 1800 |
| Pointing 6 | 2021-09-18 | 4860 | 22:35:50.39 | +33:58:06.1 | 1391.64 | 6 | 1800 |
| Pointing 7 | 2021-09-20 | 4860 | 22:35:57.31 | +33:58:06.1 | 1391.64 | 6 | 1800 |
| Pointing 8 | 2021-09-24 | 4860 | 22:36:04.23 | +33:58:06.1 | 1391.64 | 6 | 1800 |
| Pointing 9 | 2021-09-26 | 4860 | 22:35:43.37 | +33:56:51.5 | 1391.64 | 6 | 1800 |
| Pointing 10 | 2021-09-27 | 4860 | 22:35:50.39 | +33:56:51.5 | 1391.64 | 6 | 1800 |
| Pointing 11 | 2021-09-29 | 4860 | 22:35:57.31 | +33:56:51.5 | 1391.64 | 6 | 1800 |
| Pointing 12 | 2021-09-28 | 4860 | 22:36:04.23 | +33:56:51.5 | 1391.64 | 6 | 1800 |
| Pointing 13 | 2021-09-30 | 4860 | 22:35:43.37 | +33:55:36.9 | 1391.64 | 6 | 1800 |
| Pointing 14 | 2021-10-03 | 4860 | 22:35:50.39 | +33:55:36.9 | 1391.64 | 6 | 1800 |
| Pointing 15 | 2021-10-01 | 4860 | 22:35:57.31 | +33:55:36.9 | 1391.64 | 6 | 1800 |
| Pointing 16 | 2021-10-05 | 4860 | 22:36:04.23 | +33:55:36.9 | 1391.64 | 6 | 1800 |

**a:** Total observation time including all overheads.

**b:** Coordinates of the pointing position of the central beam M01.

**c:** Total on-target time.

**d:** These earlier observations have higher overheads than the later observations.

**Extended Data Table 2 | Properties of beams in the 19-beam receiver**

| Beam ID | half-power size [a] [arcmin] | gain factor [b] (Jy/K) | rms [c] (mJy beam$^{-1}$) | δ rms [d] (mJy beam$^{-1}$) |
|---|---|---|---|---|
| M01 | 2.8335 | 16.0128 | 0.18 | 0.07 |
| M02 | 2.8635 | 15.0520 | 0.14 | 0.03 |
| M03 | 2.8535 | 15.5324 | 0.13 | 0.02 |
| M04 | 2.8780 | 15.2122 | 0.16 | 0.04 |
| M05 | 2.8480 | 15.6638 | 0.15 | 0.04 |
| M06 | 2.9680 | 15.0520 | 0.16 | 0.04 |
| M07 | 2.8690 | 14.9207 | 0.16 | 0.04 |
| M08 | 2.9090 | 14.3089 | 0.17 | 0.03 |
| M09 | 2.8935 | 14.3089 | 0.16 | 0.04 |
| M10 | 2.9590 | 15.2409 | 0.14 | 0.03 |
| M11 | 2.9290 | 14.6004 | 0.14 | 0.03 |
| M12 | 2.9590 | 14.2802 | 0.15 | 0.03 |
| M13 | 2.9190 | 15.2409 | 0.14 | 0.03 |
| M14 | 2.9635 | 14.6004 | 0.14 | 0.03 |
| M15 | 2.8990 | 13.9599 | 0.16 | 0.02 |
| M16 | 2.9790 | 13.3194 | 0.26 | 0.14 |
| M17 | 2.9290 | 13.8285 | 0.17 | 0.05 |
| M18 | 2.9545 | 13.7710 | 0.17 | 0.04 |
| M19 | 2.9190 | 14.2802 | 0.15 | 0.03 |

**a:** Half-power beam size at 1391 MHz.

**b:** Gain factor at 1391 MHz.

**c:** Mean rms of the 16 spectra (bin-width of 20 km s$^{-1}$) obtained with a given beam.

**d:** Standard deviation of the rms.

**Extended Data Table 3 | Two new HI sources in the SQ neighborhood**

| Source ID | R.A. (2000J) | Decl. (2000J) | $v_{HI}$ (km s$^{-1}$) | $\Delta v_{20}$ (km s$^{-1}$) | $M_{HI}$ ($10^8 M_\odot$) |
|---|---|---|---|---|---|
| NGC 7320a | 22:36:32.2 | +33:47:46 | 6702 ± 24 | 360 | 6.3 ± 0.2 |
| Anon 7 | 22:35:19.3 | +34:08:00 | 6654 ± 16 | 160 | 2.2 ± 0.5 |